\documentclass[pre,preprint,superscriptaddress,showpacs,showkeys]{revtex4}
\usepackage{graphicx,bm,amssymb,amsmath}
\begin{document}
\title{High-frequency absorption and gain in superlattices: Semiquasistatic approach}
\author{A. V. Shorokhov}
\affiliation{Department of Physical Science, P.O. Box 3000,
University of Oulu FI-90014, Finland}
\affiliation{Institute of
Physics and Chemistry, Mordovian State University, 430000 Saransk,
Russia}
\author{K. N. Alekseev}
%%\email{}
%
\affiliation{Department of Physical Science, P.O. Box 3000,
University of Oulu FI-90014, Finland}

%\date{\today}
\begin{abstract}
We consider a generation and an amplification of THz radiation in
semiconductor superlattices under the action of microwave pump
field. Electrons belonging to a single miniband of the
superlattice interact quasistatically with the pump field and
dynamically with a signal THz field. Within this semiquasistic
approach we derive elegant difference formulas describing
absorption (gain) of the weak THz signal. We present an
instructive geometric interpretation of the absorption formulas
which allows a search of optimum conditions for the gain employing
only a simple qualitative analysis. Our theoretical findings
contribute to the development of sources and detectors of THz
radiation that are using nonlinear electric properties of
semiconductor superlattices.
\end{abstract}
\pacs{73.21.Cd, 07.57.Hm, 72.20.Ht}
\keywords{semiconductor superlattice; Terahertz
radiation; parametric generation and amplification; quasistatic interaction, quantum derivative}

\maketitle

\section{Introduction}
Semiconductor superlattices (SSLs) \cite{esa70} have attracted
growing attention in  view of their unique electronic properties,
which can be used for generation, amplification and detection of a
high-frequency electromagnetic radiation
\cite{wackerrew,platerorew,frenchrev}. A dc-biased SSL, operating
in the conditions of single miniband transport regime and
demonstrating the static negative differential conductance, can be
potentially used as an active element of THz field generator
\cite{kti72}. However, the static negative differential
conductance makes SSL unstable against a formation of high-field
electric domains \cite{But77,Ign87}. The electric domains are
believed to be destructive for the THz gain in SSLs \cite{Sav04}.
Currently the main focus is on the possibilities to overcome this
drawback within the scheme of dc-biased SSL
\cite{Sav04,Kre11,Hyart,Lis,Fei}. However, schemes of THz
superlattice devices with ac pump fields are also under discussion
\cite{Klap04,Ren05,Ale05_1,Ale05_2}. A strong microwave field
$E_p(t)=E_1\cos(\omega_1t)$ pumps the SSL and a desirable signal
field $E_s(t)=E_2\cos(\omega_2t)$ has a higher frequency
$\omega_2>\omega_1$. Because for typical SSLs the characteristic
scattering time $\tau$ at room temperature is of the order of
$100$ fs, an interaction of the microwave fields with the miniband
electrons is quasistatic ($\omega_1\tau\ll 1$). Importantly, we
have showed recently that such quasistatic pump field can
completely suppress domains in SSLs \cite{Ale05_1,Ale05_2}. Two
distinct possibilities exist for the signal field: It can be also
quasistatic ($\omega_2\tau\ll 1$) or it cannot be described within
the quasistatic approach if $\omega_2\tau\gtrsim 1$. The later
situation, i.e. $\omega_1\tau\ll 1$ but $\omega_2\tau\gtrsim 1$,
can be called \textit{semiquasistatic interaction}. Thus, the
semiquasistatic approach is introduced to describe an
amplification of THz field in SSL under the action of microwave
field. We should note that while in the experiments
\cite{Klap04,Ren05} the interaction of electrons with both pump
and signal fields is still quasistatic and, moreover, it is
probably domains-mediated, the ultimate goal of this activity is
to reach a parametric THz generation without electric domains
\cite{Ren05}. Therefore, we can say that the semiquasistatic
approach describes the real experimental situations as well.
\par
In this paper, we derive several beautiful formulas describing a
high-frequency absorption (gain) of small-signal field in the
presence of ac pump field, when the interaction of the fields with
the miniband electrons in the superlattice is semiquasistatic.
Along with obvious potential applications in the calculations of
gain in superlattice devices, our theoretical approach provides a
powerful tool for finding the correspondence between quasistatic
and dynamic regimes in ac-driven superlattices.
\par
Our main findings are following.
We consider the response of electrons, belonging to a single miniband of SSL, to an action of the
total electric field
%------Eq 1-------------------------------
\begin{equation}
\label{total_field}
E(t)=E_0+E_1\cos(\omega_1t)+E_2\cos(\omega_2t),
\end{equation}
%-----------------------------------------
where $E_0$ is the dc bias and  $E_{s}=E_2\cos(\omega_2t)$ is the weak signal (probe)
field. We should note that in the real devices $E_{s}$ is a mode of the
resonator tuned to a desirable THz frequency. Here three distinct cases should be considered:
(i) $\omega_2$ and $\omega_1$ are incommensurable, (ii) $\omega_2=m\omega_1/2$
($m$ is an odd number) and (iii) $\omega_2=m\omega_1$ ($m$ is an integer).
The later case of the parametric cascading is most interesting physically \cite{Ale05_2}
and we mainly focus on it in this paper.
We define the dimensionless absorption of the weak ac probe field in SSL as
%------Eq 2-------------------------------
\begin{equation}
\label{start}
A(\omega_2)=2\langle
V(t)\cos(\omega_2 t)\rangle_t=\frac{2}{T}\int\limits_{0}^{T}V(t)\cos(\omega_2t)\,dt,
\end{equation}
%-----------------------------------------
where $V(t)$ is the scaled miniband electron velocity defined in the units of the maximal miniband velocity
$V_0$ \cite{esa70,wackerrew}, averaging $\langle\ldots\rangle_t$ is performed over time and $T=2\pi
m/\omega_2=2\pi/\omega_1$ is the common period for the probe and the pump ac fields.
\par
Starting from the exact formal solution of the Boltzmann transport equation, we represent
the absorption $A$ as the sum of three terms
%------Eq 3-------------------------------
\begin{equation}
A=A^{harm}+A^{coh}+A^{incoh}.
\end{equation}
%-----------------------------------------
Here $A^{harm}$, $A^{coh}$ and $A^{incoh}$ describe the absorption
(gain) seeded by generation of harmonics, the parametric
amplification of the probe field due to a coherent interaction of
the pump and the probe fields, and the nonparametric absorption,
correspondingly.
\par
The term $A^{harm}$ is just the expression for $m$th in-phase
harmonic of the time-dependent current through SSL
%------Eq 4-------------------------------
\begin{equation}
\label{harm-gen}
A^{harm}=2\left\langle
I(U_{dc}+U_{ac}\cos(\omega_1t))\cos(m\omega_1t)\right\rangle_t,
\end{equation}
%-----------------------------------------
where $U_{dc}=eLE_0$ is the dc voltage, $L=Nd$ is the length of
SSL ($d$ is the period of SSL and $N$ is the number spatial
periods), $U_{ac}=eLE_1$ is the amplitude of ac voltage created by
the pump field across SSL, the current $I(t)$ is normalized to the
maximal current in SSL, $I_0$,  corresponding to the maximal
miniband velocity $V_0$. The explicit expressions for the $m$th
harmonic of the current, that determine the dependence
$A^{harm}(U_{dc},U_{ac})$, are well-known (see
\cite{esaki71,rom72} for the case of a small pump amplitude,
$U_{ac}\ll 1$, and \cite{ignatov76,pavlovich76} for the case of an
arbitrary $U_{ac}$). Note that $A^{harm}$ does not depend on the
probe; it gives the main contribution to the absorption of a weak
probe \cite{Ale05_1,Ale05_2}.
\par
The expression for $A^{harm}$ is not specific for the
semiquasistic limit because of its independence on $\omega_2$.
However, \textit{our main finding is that the absorption
components $A^{coh}$ and $A^{incoh}$ can be represented within the
semiquasistatic approach using the specific terms of quantum
derivatives} as
%------Eqs 5 -------------------------------
\begin{equation}
\label{coh}
A^{coh}=e U_2\left\langle\frac{I^{ET}\big(U_{dc}+
U_{ac}\cos(\omega_1t)+N\hbar\omega_2/e\big)-I^{ET}\big(U_{dc}+
U_{ac}\cos(\omega_1t)-N\hbar\omega_2/e\big)}{2N\hbar\omega_2}\cos(2m\omega_1t)\right\rangle_t,\\
\end{equation}
%-----------------------------------------
%------Eqs 6-------------------------------
\begin{equation}
\label{inc}
A^{incoh}=e U_2\left\langle\frac{I^{ET}\big(U_{dc}+
U_{ac}\cos(\omega_1t)+N\hbar\omega_2/e\big)-I^{ET}\big(U_{dc}+
U_{ac}\cos(\omega_1t)-N\hbar\omega_2/e\big)}{2N\hbar\omega_2}\right\rangle_t,
\end{equation}
%-----------------------------------------
where $U_2$ is the amplitude of small-signal voltage and
%------Eqs 7-------------------------------
\begin{equation}
\label{volt}
I^{ET}(U)=\frac{U/U_c}{1+(U/U_c)^2}
\end{equation}
%-----------------------------------------
is the Esaki-Tsu voltage-current (UI) characteristic
($U_c=\hbar N/e\tau$ is the critical voltage and $I^{ET}$ is normalized to the maximal current
$I_0\equiv 2I^{ET}(U=U_c)\propto V_0$).
\par
Importantly, following Eq.~(\ref{inc}) in order to find the
incoherent absorption in SSL at arbitrary high frequency
$\omega_2$ we need to know only
%------Eqs 8-------------------------------
\begin{equation}
\label{VI}
I_{dc}(U_{dc})=\left\langle I^{ET}\big(U_{dc}+U_{ac}\cos(\omega_1t)\big)\right\rangle_t,
\end{equation}
%-----------------------------------------
that is, \textit{the time-averaged current induced by the quasistatic
field (voltage)}. For a given amplitude of the ac voltage $U_{ac}$, the dc current $I_{dc}$ is a
function of only dc bias $U_{dc}$.
It is easy to calculate or to measure the modifications of UI characteristic
caused by the action of microwave (quasistatic) field \cite{Sho96_1,Win97}.
\par
On the other hand, finding of the coherent component of absorption
at the high frequency, which corresponds to the $m$th harmonic of
the pump frequency ($\omega_2=m\omega_1$), is reduced following
Eq.~(\ref{coh}) to \textit{the calculation of the $2m$th harmonic
of the current within the quasistatic approach}. This result is
especially important, because the term $A^{coh}$ can be
responsible for the parametric gain in ac-driven SSLs
\cite{Ale05_2}. Thus, one needs to know
%------Eq 9-------------------------------
\begin{equation}
\label{harm-Vm}
V_m^{harm}(U_{dc})=2\left\langle
I^{ET}(U_{dc}+U_{ac}\cos(\omega_1t))\cos(m\omega_1t)\right\rangle_t.
\end{equation}
%-----------------------------------------
(Note that in comparison with the expression (\ref{harm-gen}) we
have in Eq.~(\ref{harm-Vm}): $I(U)\rightarrow I^{ET}(U)$). For a
given pump amplitude $U_{ac}$ the amplitudes of harmonics of the
quasistatic current $V_m^{harm}$ are functions of only dc bias
$U_{dc}$. We underline that quasistatic calculations of harmonics
$V_m^{harm}$ are among the standard theoretical tools to explain
experimental results on the frequency multiplications in the
microwave frequency band \cite{Gre95,Sho96_2}.
\par
In this paper we will present simple but instructive geometric
interpretations of the formulas (\ref{coh}) and (\ref{inc})
allowing to find the high-frequency absorption components directly
using only the ruler and the knowledge of the quasistatic curves
$V_m^{harm}(U_{dc})$ and $I_{dc}(U_{dc})$. These geometric
interpretations provide a powerful tool in a search of the optimum
conditions for the gain which employs only a simple qualitative
analysis.
\par
If the frequency of the probe $\omega_2$  incommensurates with the
frequency of the pump $\omega_1$, the expression for the
absorption at $\omega_2$ consists of only one term -- the
incoherent absorption $A=A^{incoh}$ with semiquasistatic limit
(\ref{inc}). Additionally, we will also show that within the
semiquasistatic approach the difference formulas similar to
Eqs.~(\ref{coh})-(\ref{inc}) can be written
 for the absorption at half-harmonics $\omega_2=m\omega_1/2$.
\par
The organization of this paper is as follows. We start with the
calculations of absorption for arbitrary pump and probe
frequencies solving the Boltzmann transport equation in
sec.~\ref{sec:general_case}. The derivation of the absorption
formulas (\ref{coh}) and (\ref{inc}) within the semiquasistatic
limit is presented in  sec.~\ref{sec:semiquasi}. In the
sec.~\ref{sec:quasi}, an additional limitation on the frequency of
the probe is imposed and the quasistatic formulas for the
absorption  are obtained. Section~\ref{sec:numerical} is devoted
to the presentation of useful geometric interpretations of the
semiquasstatic results and their numerical verifications. The
summary of the work, as well as a brief discussion of the place of
our findings among other works devoted to the use of quantum
derivatives in solid-state systems, are presented in the final
section. Appendixes \ref{app:coherent} and \ref{app:harmonic} are
devoted respectively to the derivation of the semiquasistatic
limit for the coherent component of absorption and to the
derivation of the quasistatic limit for the harmonic component of
absorption.
\section{\label{sec:general_case}Absorption in the general case}
In this section we will represent the expressions for the
components of absorption that follow from the exact formal
solution of the transport Boltzmann equation with a constant
relaxation time \cite{chambers,budd}. Using this exact solution,
we found \cite{Sho06} that for general case of commensurate
frequencies $\omega_1/\omega_2=n/m$ ($n,m$ are integers and $n/m$
is an irreducible fraction) the total absorption $A(\omega_2)$,
Eq.~(\ref{start}) has the form
%------Eq 10-------------------------------
\begin{eqnarray}
\label{absorp}
A(\omega_2)&=&\sum_{l_1,l_2=-\infty}^{\infty}\sum_{j=-\infty}^{\infty}
J_{l_1}(\beta_1)J_{l_2}(\beta_2)J_{l_1-jm}(\beta_1)\left[J_{l_2+jn-1}(\beta_2)+
J_{l_2+jn+1}(\beta_2)\right]\nonumber\\
&\times&\frac{(\Omega_0+l_1\omega_1+l_2\omega_2)\tau}
{1+(\Omega_0+l_1\omega_1+l_2\omega_2)^2\tau^2},
\end{eqnarray}
%-----------------------------------------
where $\beta_{i}=\Omega_i/\omega_i$, $\Omega_i=edE_i/\hbar$
($i=1,2$), and $\Omega_0=edE_0/\hbar$ is the Bloch frequency. To
consider the absorption at $m$th harmonic of the pump frequency
one should fix $n=1$; for the case of absorption at half-harmonics
$n$ should be equal to $2$.
\par
First, we consider the absorption at harmonics of the pump field.
In the limit of the weak probe ($\beta_2\ll 1$), we need to take
only certain combinations of the Bessel functions indexes. As a
result we represent the absorption as a sum of three terms
\cite{Sho06}
%------Eq 11-------------------------------
\begin{equation}
A=A^{harm}+A^{coh}+A^{incoh}+O(\beta_2^2),
\end{equation}
%-----------------------------------------
%------Eq 12-------------------------------
\begin{eqnarray}
\label{Aharm}
A^{harm}&=&\sum_{l=-\infty}^{\infty}J_{l}(\beta_1)\left[J_{l-m}(\beta_1)
+J_{l+m}(\beta_1)\right]\frac{(\Omega_0+l\omega_1)\tau}
{1+(\Omega_0+l\omega_1)^2\tau^2}
\end{eqnarray}
%-----------------------------------------
%------Eq 13-------------------------------
\begin{eqnarray}
\label{Acoh}
A^{coh}&=&\frac{\beta_2}{2}\sum_{l=-\infty}^{\infty}J_{l}(\beta_1)\left[J_{l-2m}(\beta_1)
-J_{l+2m}(\beta_1)\right]\frac{(\Omega_0+l\omega_1)\tau}
{1+(\Omega_0+l\omega_1)^2\tau^2},
\end{eqnarray}
%-----------------------------------------
%------Eq 14-------------------------------
\begin{eqnarray}
\label{Aincoh}
A^{incoh}&=&\frac{\beta_2}{2}\sum_{l=-\infty}^{\infty}J_{l}^2(\beta_1)
\left[\frac{(\Omega_0+l\omega_1+\omega_2)\tau}
{1+(\Omega_0+l\omega_1+\omega_2)^2\tau^2}-\frac{(\Omega_0+l\omega_1-\omega_2)\tau}
{1+(\Omega_0+l\omega_1-\omega_2)^2\tau^2}\right].
\end{eqnarray}
%-----------------------------------------
\par
Second, for the absorption at half-harmonics of the pump
frequency we get from (\ref{absorp}) the expression for $A(\omega_2)$
as the sum of two terms
%------Eq 15-------------------------------
\begin{equation}
\label{half}
A_h=A_h^{coh}+A^{incoh}+O(\beta_2^2),
\end{equation}
%-----------------------------------------
where
%------Eq 16-------------------------------
\begin{eqnarray}
\label{Acohhalf}
A^{coh}_{h}&=&\frac{\beta_2}{2}\sum_{l=-\infty}^{\infty}J_{l}(\beta_1)\left[J_{l-m}(\beta_1)
-J_{l+m}(\beta_1)\right]\frac{(\Omega_0+l\omega_1)\tau}
{1+(\Omega_0+l\omega_1)^2\tau^2}
\end{eqnarray}
%-----------------------------------------
and $A^{incoh}$ is given by the formula (\ref{Aincoh}).
\par
Finally, it can be shown that if $\omega_2/\omega_1$ is some
irrational number, then the absorption is completely determined by
the incoherent component: $A(\omega_2)=A^{incoh}$.
\par
Now we turn to the consideration of semiquasistatic limits of
different absorption components.
\section{\label{sec:semiquasi}Derivation of the semiquasistatic formulas}
The derivation of semiquasistatic formulas is based on the use of the following
integral representation of the Esaki-Tsu characteristic \cite{esa70}
%------Eq 17-------------------------------
\begin{equation}
\label{ET_integr}
I^{ET}(\omega)=\frac{\tau\omega}{1+\tau^2\omega^2}\equiv\frac{1}{\tau}
\int\limits_{0}^{\infty}\exp\left(-\frac{t}{\tau}\right)\sin(\omega t) dt.
\end{equation}
%-----------------------------------------
and the asymptotic saddle-point method \cite{Fed77}.
\par
We start with the consideration of semiquasistatic limit for the
term $A^{incoh}$, Eq.~(\ref{Aincoh}). We rewrite $A^{incoh}$ in
terms of $I^{ET}$ (\ref{ET_integr}) as
%------no number-------------------------------
\begin{eqnarray}
%\label{Aexp}
%
&&A^{incoh}=\frac{\beta_2}{2}\sum_{l=-\infty}^{\infty}J_{l}^2(\beta_1)
\left[I^{ET}(\Omega_0+l\omega_1+\omega_2)-I^{ET}(\Omega_0+l\omega_1-\omega_2)\right]
\nonumber\\&=&\frac{\beta_2}{2\tau}\int\limits_{0}^{\infty}\exp\left(-\frac{t}{\tau}\right)
\sum_{l=-\infty}^{\infty}J_{l}^2(\beta_1)
\Big\{\sin\big[(\Omega_0+l\omega_1+\omega_2)t\big]
-\sin\big[(\Omega_0+l\omega_1-\omega_2)t\big]\Big\}\,dt.\nonumber
\end{eqnarray}
%-----------------------------------------
Using the addition formula
%------Eq no number-------------------------------
$$
\sin\big[(\Omega_0+l\omega_1+\omega_2)t\big]=\sin\big[(\Omega_0+\omega_2)t\big]\cos(l\omega_1t)+
\cos\big[(\Omega_0+\omega_2)t\big]\sin(l\omega_1t)
$$
%-----------------------------------------
and taking into account that the sum with $\sin(l\omega_1t)$ is
equal to zero, we get
%------Eq 18-------------------------------
\begin{eqnarray}
\label{promegut1}
&&A^{incoh}=\frac{\beta_2}{2\tau}\int\limits_{0}^{\infty}\exp\left(-\frac{t}{\tau}\right)\sum_{l=-\infty}^{\infty}J_{l}^2(\beta_1)
\Big\{\sin\big[(\Omega_0+l\omega_1+\omega_2)t\big]\nonumber\\&-&\sin\big[(\Omega_0+l\omega_1-\omega_2)t\big]\Big\}\,dt
=\frac{\beta_2}{2\tau}\int\limits_{0}^{\infty}\exp\left(-\frac{t}{\tau}\right)J_{0}\left(2\beta_1\sin\frac{\omega_1t}{2}\right)\nonumber\\
&\times& \Big\{\sin\big[(\Omega_0+\omega_2)t\big]
-\sin\big[(\Omega_0-\omega_2)t\big]\Big\}\,dt.
\end{eqnarray}
%-----------------------------------------
In Eq.~(\ref{promegut1}) we also have used the equality
\cite{Pru90}
%------Eq no number-------------------------------
$$
\sum_{l=-\infty}^{\infty}J_{l}^2(\beta_1)\cos(l\omega_1t)=
J_0\left(2\beta_1\sin\frac{\omega_1t}{2}\right).
$$
%-----------------------------------------
As a next step, we substitute the integral representation for $J_0(z)$ \cite{Pru90}
%------Eq no number-------------------------------
$$
J_0(z)=\frac{1}{\pi}\int\limits_{0}^{\pi}\cos(z\cos\Theta)\,d\Theta
$$
%-----------------------------------------
in (\ref{promegut1}) and get
%------Eq 19-------------------------------
\begin{eqnarray}
\label{sumin}
&&A^{incoh}=\frac{\beta_2}{2\pi\tau}\int\limits_{0}^{\infty}
\exp\left(-\frac{t}{\tau}\right)\int\limits_{0}^{\pi}
\cos\left(2\beta_1\sin\frac{\omega_1t}{2}\cos\Theta\right)\,d\Theta
\Big\{\sin\big[(\Omega_0+\omega_2)t\big]\nonumber\\&-&
\sin\big[(\Omega_0-\omega_2)t\big]\Big\}\,dt=
\frac{\beta_2}{2\pi\tau}\int\limits_{0}^{\infty}\exp\left(-\frac{t}{\tau}\right)
\int\limits_{0}^{\pi}\left\{\sin\left[\left(\Omega_1\frac{\sin(\omega_1t/2)}
{\omega_1t/2}\cos\Theta+\Omega_0+\omega_2\right)t\right]\right.\nonumber\\
&-&\left.\sin\left[\left(\Omega_1\frac{\sin(\omega_1t/2)}
{\omega_1t/2}\cos\Theta+\Omega_0-\omega_2\right)t\right]\right\}\,d\Theta\,
dt \stackrel{\omega_1\tau\ll1}{\longrightarrow} \nonumber\\
&
&\frac{\beta_2}{4\pi\tau}\int\limits_{0}^{\infty}\exp\left(-\frac{t}{\tau}\right)
\int\limits_{0}^{2\pi}\Big\{\sin\left[\left(\Omega_1
\cos\Theta+\Omega_0+\omega_2\right)t\right]-\sin\left[\left(\Omega_1
\cos\Theta+\Omega_0-\omega_2\right)t\right]\Big\}\, d\Theta
\, dt . \nonumber\\
\end{eqnarray}
%----------------------------------------------------
In (\ref{sumin}) we first used  simple formula
$2\cos\alpha\sin\beta=\sin(\alpha+\beta)-\sin(\alpha-\beta)$ and
then took into account that for $\omega_1\tau\ll 1$ the main
contribution to the integrals with the exponential factor
$\exp(-t/\tau)=\exp(-\omega_1 t/\omega_1 \tau)$ comes from the
terms formally satisfying $\omega_1 t\rightarrow 0$ (the
saddle-point method). Finally, using Eq.~(\ref{ET_integr}) we make
the inverse transformation to the Esaki-Tsu current and obtain
%------Eq 20-------------------------------
\begin{equation}
\label{incohsemi}
A^{incoh}=\frac{\beta_2}{4\pi}\int\limits_{0}^{2\pi}\left[I^{ET}\left(\Omega_1
\cos\Theta+\Omega_0+\omega_2\right)-I^{ET}\left(\Omega_1
\cos\Theta+\Omega_0-\omega_2\right)\right]\,d\Theta.
\end{equation}
%----------------------------------------------------
Now it is easy to see that after making the transformation from
frequency to  voltage variables and using the Esaki-Tsu
characteristic in the form (\ref{volt}), we get for $A^{incoh}$
the formula (\ref{inc}).
\par
Applying the same method, one can obtain (see Appendix~\ref{app:coherent}) in the
semiquasistatic limit the following expression for the coherent component of absorption
%------Eq 21-------------------------------
\begin{equation}
\label{cohsemi}
A^{coh}=\frac{\beta_2}{4\pi}\int\limits_{0}^{2\pi}
\left[I^{ET}(\Omega_1\cos\Theta+\Omega_0+\omega_2)-
I^{ET}(\Omega_1\cos\Theta+\Omega_0-\omega_2)\right]\cos(2m\Theta)\,d\Theta,
\end{equation}
%----------------------------------------------------
which is equivalent to the formula (\ref{coh}).
\par
We turn to the discussion of the semiquasistatic limit
for absorption in the cases of incommensurable frequencies and half-harmonics.
As we have found in the previous section, the absorption for incommensurable frequencies
is just $A^{incoh}$ defined by Eq.~(\ref{Aincoh}).
Therefore, its semiquasistatic limit is just the formula (\ref{incohsemi}).
\par
Next, in the general case of arbitrary values of $\omega_1\tau$
and $\omega_2\tau$, the coherent component of absorption at
half-harmonics $A_h^{coh}$ has the same form as the coherent
component of absorption at harmonics (\ref{Acoh}), if one makes a
formal change $2m\rightarrow m$ (cf. Eqs. (\ref{Acohhalf}) and
(\ref{Acoh})). Therefore, the quasistatic limit of $A_h^{coh}$
coincides with Eq.~(\ref{coh}) after the substitution
$2m\rightarrow m$. However, we should note that the consistent
proof of this statement is not trivial because the derivations of
semiquasistatic formulas for odd and for even values $m$ have
different ways (in the case of half-harmonics $m$ is the odd
number but in the case of harmonics $2m$ is the even number).
\section{\label{sec:quasi}Quasistatic limit}
To find the quasistatic limit of the components of absorption we
have to impose the additional limitation on $\omega_2$:
$\omega_2\ll \tau^{-1}$. For $A^{harm}$ we obtain from
Eq.~(\ref{Aharm}) in the quasistatic limit
%------Eq 22-------------------------------
%\begin{eqnarray}
\begin{equation}
\label{harmquasi}
A^{harm}=\frac{1}{\pi}\int\limits_{0}^{2\pi}
I^{ET}(\Omega_0+\Omega_1\cos\Theta)\cos(m\Theta)\, d\Theta.
%\end{eqnarray}
\end{equation}
%-------------------------------------------
Of course, this expression is identical to Eq.~(\ref{harm-Vm}). Formally,
Eq.~(\ref{harmquasi}) follows from (\ref{harm-gen}) after the
substitution $I(U)\rightarrow I^{ET}(U)$, but the correct and
consistent derivation of this result still requires some algebra
(see Appendix~\ref{app:harmonic}).
\par
To get the quasistatic limit for $A^{incoh}$ and $A^{coh}$ we need
to consider the limit $\omega_2\tau\ll1$ in the Eqs. (\ref{incohsemi}) and
(\ref{cohsemi}). In this limit the finite difference
$I^{ET}(\Omega_1\cos\Theta+\Omega_0+\omega_2)-
I^{ET}(\Omega_1\cos\Theta+\Omega_0-\omega_2)$ goes into the
derivative
%------Eq no number-------------------------------
%\begin{eqnarray}
$$
%%&&
\frac{I^{ET}(\Omega_1\cos\Theta+\Omega_0+\omega_2)-
I^{ET}(\Omega_1\cos\Theta+\Omega_0-\omega_2)}{2\omega_2}\rightarrow\frac{\partial
I^{ET}(\Omega_1\cos\Theta+\Omega_0)}{\partial\Omega_0}
$$
%\end{eqnarray}
%-------------------------------------------
and hence $A^{coh}$ and $A^{incoh}$ become
%------no number-------------------------------
$$
A^{coh}=\frac{\Omega_2}{2\pi}\int\limits_{0}^{2\pi}\frac{\partial
I^{ET}(\Omega_1\cos\Theta+\Omega_0)}{\partial\Omega_0}\cos(2m\Theta)\,d\Theta,
\nonumber\quad
A^{incoh}=\frac{\Omega_2}{2\pi}\int\limits_{0}^{2\pi}\frac{\partial
I^{ET}(\Omega_1\cos\Theta+\Omega_0)}{\partial\Omega_0}\,d\Theta
.\nonumber
$$
%-------------------------------------------
These formulas in  terms of voltage and current, Eq.~(\ref{volt}),
have the following form
%------Eq 23-------------------------------
\begin{eqnarray}
A^{coh}=U_2\frac{\partial }{\partial U_{dc}}\left\langle
I^{ET}\left( U_{dc}+U_{ac}\cos(\omega_1t)\right) \cos(2m\omega_1t)\right\rangle_t ,
\end{eqnarray}
%-------------------------------------------
%------Eq 24-------------------------------
\begin{eqnarray}
\label{dif} A^{incoh}=U_2\frac{\partial }{\partial
U_{dc}}\left\langle
I^{ET}(U_{dc}+U_{ac}\cos(\omega_1t))\right\rangle_t.
\end{eqnarray}
%-------------------------------------------
Thus, in the quasistatic limit the incoherent and the coherent components of absorption
at $m$th harmonic
are proportional respectively to the derivative of dc component (\ref{VI}) and to the derivative
of $2m$th harmonic (\ref{harm-Vm}).
\par
On the other hand, the formulas for the absorption components can
be derived completely within the quasistatic approach. We define
the absorption of quasistatic field (voltage) as
$A^{qst}(\omega_2)=\langle I_{ET}(U)\cos(\omega_2 t)\rangle_t$,
where the Esaki-Tsu UI characteristic $I_{ET}(U)$ is given by
Eq.~(\ref{volt}). For a weak signal we expand the current in the
Taylor series $I_{ET}(U)\approx I_{ET}(U_{p})+I_{ET}'(U_{p})\times
U_{s}$ (here prime means the derivation with respect to $U$,
$U_p=U_{dc}+ U_{ac}\cos(\omega_1t)$, $U_{s}=U_2\cos(\omega_2t)$).
Substituting this expansion to the definition of $A^{qst}$ we have
%------Eq 25-------------------------------
\begin{equation}
\label{Aesaki}
A^{qst}(\omega_2)=2\langle
I_{ET}(U_{p})\cos(\omega_2t)\rangle_t+ \langle
I_{ET}'(U_{p})\cos(2\omega_2t)\rangle_tU_{2}+ \langle I_{ET}'(U_{p})\rangle_tU_{2}
\end{equation}
%-------------------------------------------
Making comparison of Eq.~(\ref{Aesaki}) with other formulas from
this section, we see that in the limit $\omega_2\tau\rightarrow 0$
the semiquasistatic formulas for the components of absorption
correctly reproduce the quasistatic results.
\section{\label{sec:numerical}Geometrical interpretation and numerical verification}
%
%=======================Fig 1=================================
%%former {fig:1}{fig1.eps}
\begin{figure}
\includegraphics[clip=true,width=0.7\linewidth]{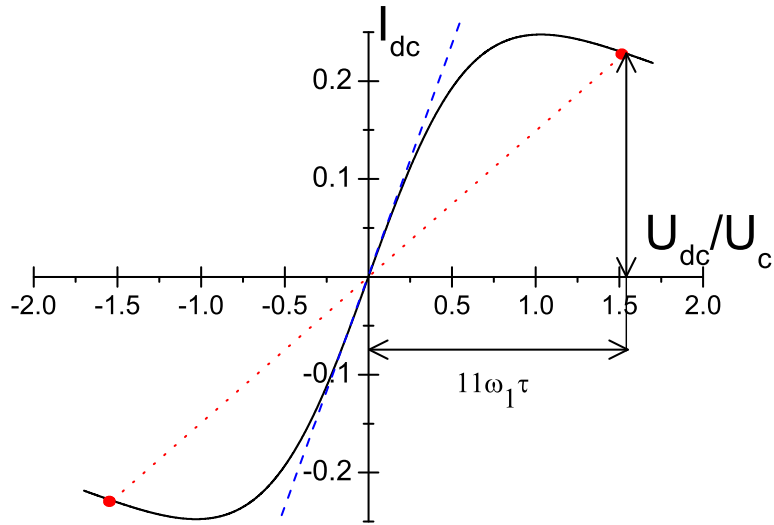}%
\caption{\label{fig:1}[Color online] Geometrical meaning of the
incoherent component of absorption within semiquasistatic and
quasistatic approaches. Time-averaged current $I_{dc}$ under the
action of quasistatic pump ($\omega_1\tau=0.1$, $U_{ac}/U_c=0.2$)
vs dc voltage $U_{dc}$. If we choose the working point at
$U_{dc}=0$, then the dotted segment [red online] corresponds to
the finite difference (quantum derivative) of a weak probe at
$\omega_2=11\omega_1$ and the dashed straight line [blue online]
corresponds to the derivative. The slopes of these strait lines
determines $A^{incoh}$ respectively in  semiquasistatic and
quasistatic cases.}
\end{figure}
%=======================================================================
%=======================Fig 2=============================
%%former {fig:2}{fig2_new.eps}
\begin{figure}
\includegraphics[clip=true,width=0.7\linewidth]{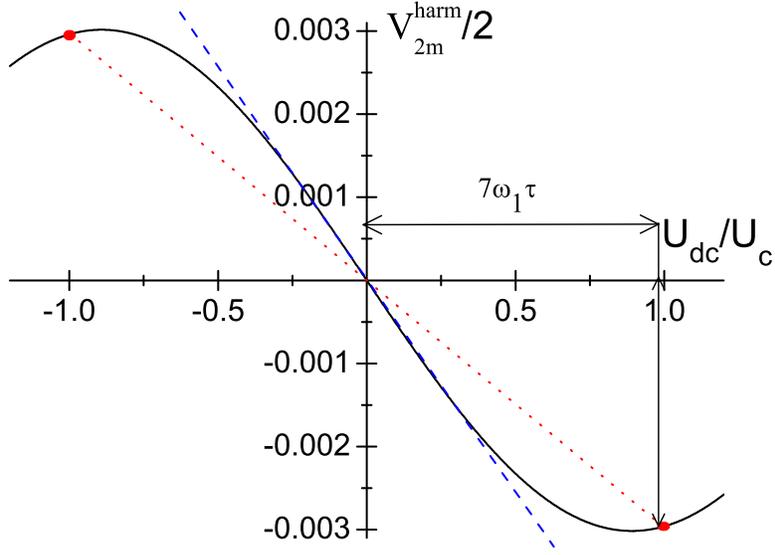}%
\caption{\label{fig:2} [color online] Geometrical meaning of the
coherent component of absorption $A^{coh}$ within semiquasistatic
and quasistatic approaches. The $2m$th harmonic of quasistatic
current $V_{2m}^{harm}$ vs dc voltage $U_{dc}$ for $m=7$. The
current is induced by the quasistatic pump ($\omega_1\tau=0.14$)
with the amplitude $U_{ac}/U_c=8$. Slope of dotted segment [red
online] determines the (negative) finite difference for a weak
probe with $\omega_2=7\omega_1$. Slope of dashed straight line
[blue online] determines the derivative. Working point is chosen
at $U_{dc}=0$.}
\end{figure}
%=======================================================================
%==============Fig 3 ===================================================
%%former {fig:10}{fig10.eps}
\begin{figure}
\includegraphics[clip=true,width=0.7\linewidth]{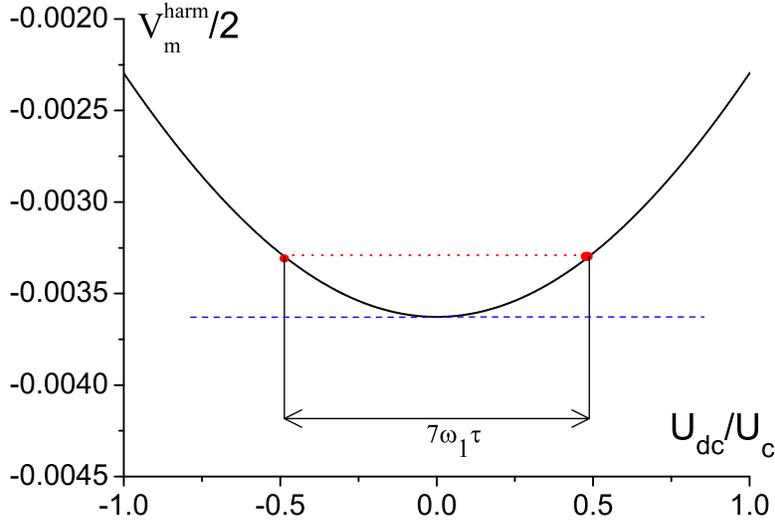}%
\caption{\label{fig:3} [color online] Geometric interpretation for
the coherent component of absorption at half-harmonics $A^{coh}_h$
within semiquasistatic (dotted [red online] segment) and
quasistatic (dashed [blue online] line). The curve represents
$m$th harmonic of quasistatic current $V_{m}^{harm}$ vs dc voltage
$U_{dc}$ for $\omega_2=7\omega_1/2$, $\omega_1\tau=0.14$ and
$\Omega_1\tau=8$. The zero slopes of the straight lines show that
$A^{coh}_h$ calculated for $U_{dc}=0$ is zero.}
\end{figure}
%================================================================
%=======================Fig 4=================================
%%former {fig:3}{fig3.eps}
\begin{figure}
\includegraphics[clip=true,width=0.7\linewidth]{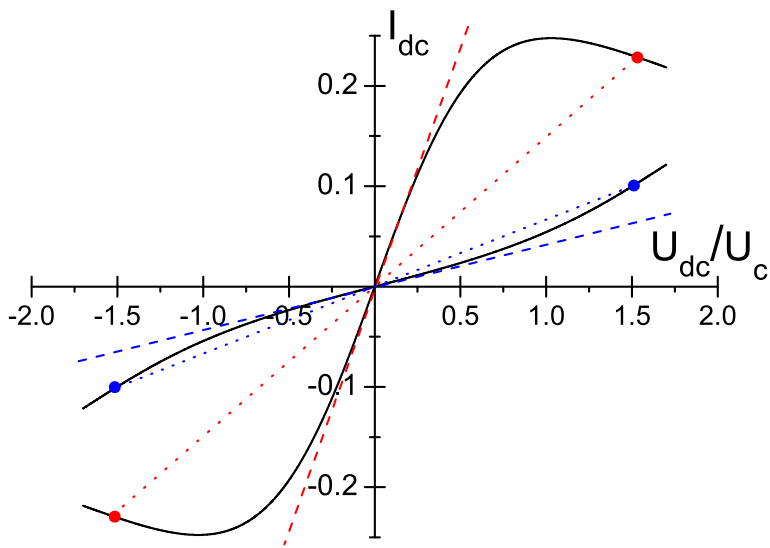}%
\caption{\label{fig:4} [Color online] Incoherent component of
absorption for different strengths of ac pump. Both difference and
usual derivatives are simultaneously small and approach each other
for the strong pump ($U_{ac}/U_c=2$) [blue online], in contrast to
the case of weak pump ($U_{ac}/U_c=0.2$) [red online] where they
are quite different. Here $\omega_1\tau=0.14$ and $m=11$.}
\end{figure}
%===============================================================
%=======================Fig 5=================================
%%former {fig:4_ins}{fig4_ins.eps}
\begin{figure}
\includegraphics[clip=true,width=0.7\linewidth]{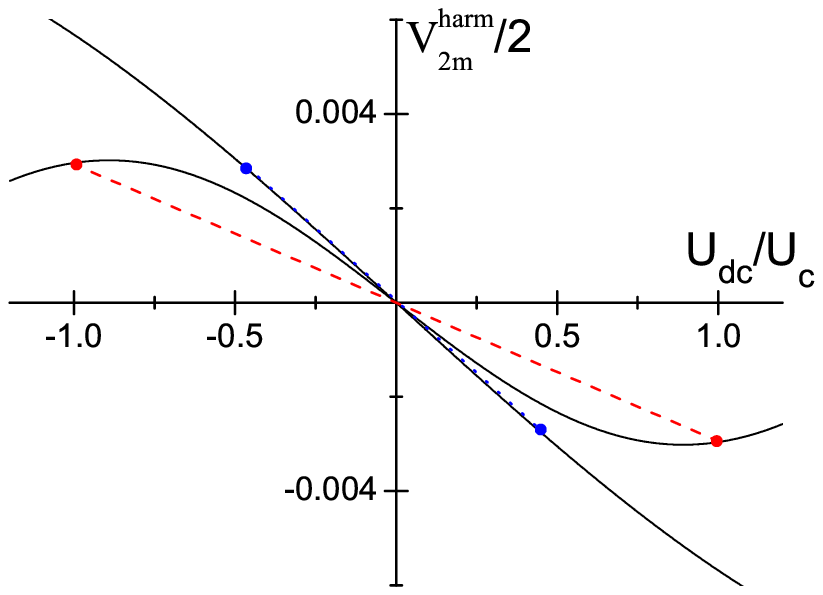}%
\caption{\label{fig:5} [Color online] Comparison of the coherent
absorptions for different values of $m=\omega_2/\omega_1$. Quantum
derivative [blue online] approaches normal derivative (the tangent
line, not shown) for $m=3$; they are quite different for $m=7$
[red online]. Quasistatic pump with $\omega\tau=0.14$ has the
amplitude $U_{ac}/U_c=8$.}
\end{figure}
%=======================================================================
To discuss the geometric meaning of incoherent and coherent
components of absorption it is convenient to present $A^{incoh}$
and $A^{coh}$ explicitly in terms of zero and even harmonics of
the quasistatic current $I^{ET}$. Combining  (\ref{coh}) and
(\ref{inc}) with (\ref{harm-Vm}) and (\ref{VI}) we have
%------Eq 26-------------------------------
\begin{equation}
\label{inc_most}
A^{incoh}=\frac{e U_2}{2\omega_2}\left[ I_{dc}\left( U_{dc}+N\hbar\omega_2/e\right)-
I_{dc}\left( U_{dc}-N\hbar\omega_2/e\right)\right]
\end{equation}
%-----------------------------------------
%------Eq 27-------------------------------
\begin{equation}
\label{coh_most}
A^{coh}=\frac{1}{2}\frac{e U_2}{2\omega_2}\left[
V_{2m}^{harm}\left( U_{dc}+N\hbar\omega_2/e\right)-
V_{2m}^{harm}\left( U_{dc}-N\hbar\omega_2/e\right)\right]
\end{equation}
%-----------------------------------------
Alternatively, the expressions for the absorption components in
terms of frequency (\ref{incohsemi}), (\ref{cohsemi}) can be used.
\par
As is evident from Fig.~\ref{fig:1}, the geometric representation
of the quantum derivative (finite difference) of $I_{dc}$ is the
segment with the length defined by the frequency of probe field
$\omega_2$. The ends of the segment belong to the curve
$I_{dc}(U_{dc})$ and their locations are determined by the
location of working point. Slope of the segment determines
$A^{incoh}$ within the semiquasistatic approach. In the
quasistatic limit, a finite difference becomes an usual derivative
and therefore $A^{incoh}$ is determined just by the slope of the
tangent line to $I_{dc}(U_{dc})$ curve at the working point. In a
similar manner, the coherent component of absorption $A^{coh}$ at
the frequency $\omega_2=m\omega_1$ is determined by the slope of
the segment with the ends  belonging to the curve
$V_{2m}^{harm}(U_{dc})$ (Fig.~\ref{fig:2}). Respectively, the
slope of the tangent line to $V_{2m}^{harm}(U_{dc})$ at the
working point determines $A^{coh}$ in the quasistatic limit.
\par
\begin{figure}
\includegraphics[clip=true,width=0.7\linewidth]{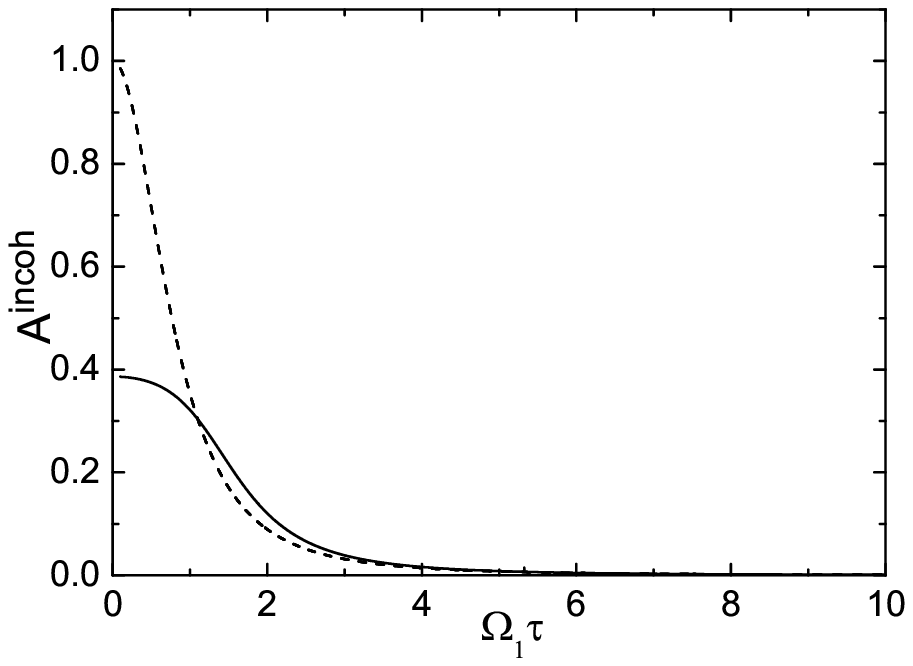}%
\caption{\label{fig:6}Dependence of $A^{incoh}$ on the amplitude
of ac pump calculated within semiquasistatic (solid) and
quasistatic (dashed) approaches for $m=9$ and
$\omega_1\tau=0.14$.}
\end{figure}

Using these two geometrical pictures it is possible, in principle,
to optimize a location of the working point (i.e. the value of
applied dc bias) in order to obtain the maximal small-signal gain
in SSL in the conditions of suppressed domains, when both
$m=\omega_2/\omega_1$ and the pump amplitude $U_{ac}$ are given.
However, such an analysis goes beyond the scope of the present
paper. In the rest of this section we will focus on the case of
dc-unbaised SSL ($U_{dc}=0$), which corresponds to the choice of
working point at the origin of coordinates in Figs.~\ref{fig:1}
and \ref{fig:2}. Such a choice guarantees the suppression of
electric domains \cite{Ale05_2}: a slope of the dependence
$I_{dc}$ on $U_{dc}$ is positive practically for all $U_{ac}$ in
the limit $U_{dc}\rightarrow 0$ (for examples, see
Figs.~\ref{fig:1}, \ref{fig:4}, \ref{fig:6}).
\par
To illustrate the usefulness of the geometric interpretation in
finding of schemes allowing gain in the conditions of suppressed
domains, we will consider two characteristic examples (earlier the
solutions of these problems have been announced without proofs in
\cite{Ale05_2}).
\par
\textit{Problem I}. Suppose that $\omega_2/\omega_1$ is an
irrational number and therefore only the incoherent absorption
$A^{incoh}$ contributes to the gain in SSL. Employing the
geometrical interpretation of $A^{incoh}$, we immediately see that
if the slope of the tangent line to $I_{dc}$ curve is positive,
then the quantum derivative is also positive for an arbitrary
$\omega_2$ (see, e. g., Figs.~\ref{fig:1},\ref{fig:4}). In
physical terms, this means that a small-signal gain in the
unbiased SSL is impossible in the conditions of suppressed
domains, if the frequencies of pump and probe are incommensurable.
\par
\textit{Problem II}. Consider the absorption at a probe frequency
that is a half-harmonic of the pump, $\omega_2/\omega_1=m/2$. The
total absorption is the sum of the incoherent component
$A^{incoh}$ and the coherent absorption at half-harmonics
$A^{coh}_h$ (see Eq.~\ref{Acohhalf}). The geometric interpretation
of $A^{coh}_h$ is similar to the interpretation of $A^{coh}$: One
needs only to consider  $V_{m}^{harm}$ as a function of $U_{dc}$
instead of a dependence $V_{2m}^{harm}(U_{dc})$. Obviously, the
dependence  of any harmonic of the quasistatic current on the
applied dc bias, i. e. $V_{m}^{harm}(U_{dc})$, is always symmetric
about the vertical line $U_{dc}=0$ Therefore, the choice of the
working point at $U_{dc}=0$ results in zero value of both the
derivative and the quantum derivative in this point
(Fig.~\ref{fig:3}). Thus, in the case of dc-unbiased SSL
$A^{coh}_h$ is always zero. On the other hand, as we have saw
earlier, $A^{incoh}$ is always positive in the conditions of
suppressed domains. Summing up we can conclude that a small-signal
gain in unbiased SSL is impossible in the conditions of suppressed
domains, if the frequency of probe is a half-harmonic of the pump
frequency.
%=======fig 6 ================================================
%%former{fig:4}{fig4.eps}
%======================================================================
%=================Fig 7 =================================
%%former{fig:5}{fig5.eps}
\begin{figure}
\includegraphics[clip=true,width=0.7\linewidth]{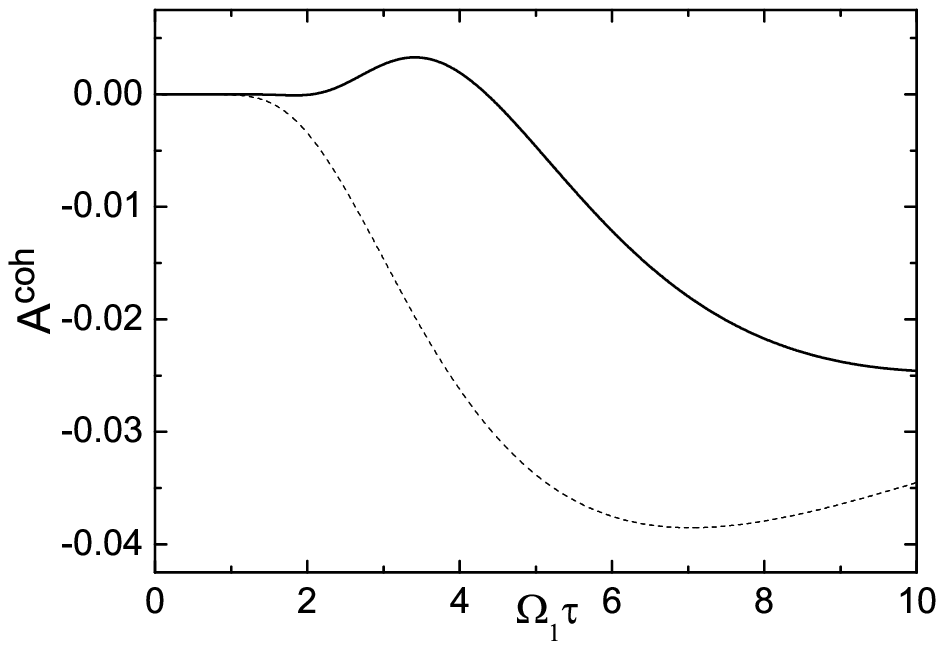}%
\caption{\label{fig:7} Same as in Fig.~\ref{fig:6} but for
$A^{coh}$ and $m=7$}.
\end{figure}
%================================================================
\par
These two conclusive findings allow to narrow the class of
possible schemes for the superlattice THz emitters with ac-pump,
which can operate in the domainless regimes. The proper scheme
should exploit a negative absorption at some particular harmonics
of the probe, when negative values of harmonic ($A^{harm}$) and
coherent contributions to the net gain can overcome always
positive incoherent absorption component \cite{Ale05_2}.
Therefore, in our analysis we return to the case
$\omega_2/\omega_1=$integer and compare $A^{incoh}$ and $A^{coh}$
in conditions of quasistatic, semiquasistatic and dynamic
approaches.
\par
Semiquasistatic and quasistatic results have a visible difference
for higher harmonics as is evident from Figs.~\ref{fig:1} and
\ref{fig:2} (here the quasistatic condition $\omega_2\tau\ll 1$ is
not satisfied for the probe field). However, in the case of
low-order harmonics the difference between semiquasistatic and
quasistatic results is very small. This fact is illustrated in
Fig.~\ref{fig:5} for $A^{coh}$, where the difference between the
derivative (tangent line) and the quantum derivative is very small
for $m=3$,  but it visible increases for $m=7$.
\par
The dependencies of $A^{incoh}$ and $A^{coh}$ on the amplitude of
pump field, calculated using semiquasistatic (solid) and
quasistatic (dashed) formulas for large $m$, are shown in
Figs.~\ref{fig:6} and \ref{fig:7}. Here $A^{incoh}$ is always
positive and its value decreases with an increase of the pump
amplitude. Fig.~\ref{fig:6} demonstrates that the quasistatic
calculations significantly overestimate $A^{incoh}$ for a small
pump $U_{ac}/U_c<1$ (see also Fig.~\ref{fig:1}). However, the
agreement between semiquasistatic and quasistatic approaches in
finding $A^{incoh}$ becomes good for a strong pump
(Fig.~\ref{fig:6}). Such paradoxical behavior is explained by the
effect of shift of the maximum of UI characteristic under the
action of quasistatic ac field \cite{Mensah_JPConM92,EPJB04}
(Note: in our case the pump is still quasistatic). As we see in
Fig.~\ref{fig:4}, for a large pump UI characteristic is rather
closed to the linear in the interval of dc voltages that is
utilized for the calculation of quantum derivative,  what is in a
sharp contrast to the case of large ac pump where UI
characteristic demonstrates a strongly nonlinear behavior in the
same interval. Thus, for a strong pump the finite difference is
close to the tangent line, providing a small difference between
the results of semiquastatic and quasistatic approaches for
$A^{incoh}$. As for $A^{coh}$, the difference between
semiquastatic and quasistatic calculations is quite visible for
large $m$ (Fig.~\ref{fig:7}). Moreover, the quasistatic approach
not only overestimate gain in this case, it can give even a wrong
sign of $A^{coh}$ in some interval of ac pump amplitudes
(Fig.~\ref{fig:7}).
\par
\begin{figure}
\includegraphics[clip=true,width=0.7\linewidth]{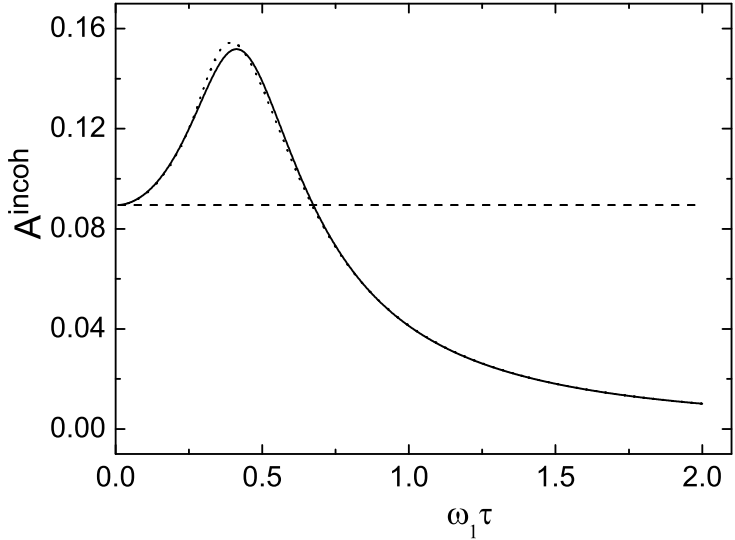}%
\caption{\label{fig:8} Dependence of $A^{incoh}$ on $\omega_1\tau$
calculated within semiquasistatic (solid), dynamic (dotted) and
quasistatic (dashed) approaches for $m=5$ and pump amplitude
$\Omega_1\tau=2.0$.}
\end{figure}
\begin{figure}
\includegraphics[clip=true,width=0.7\linewidth]{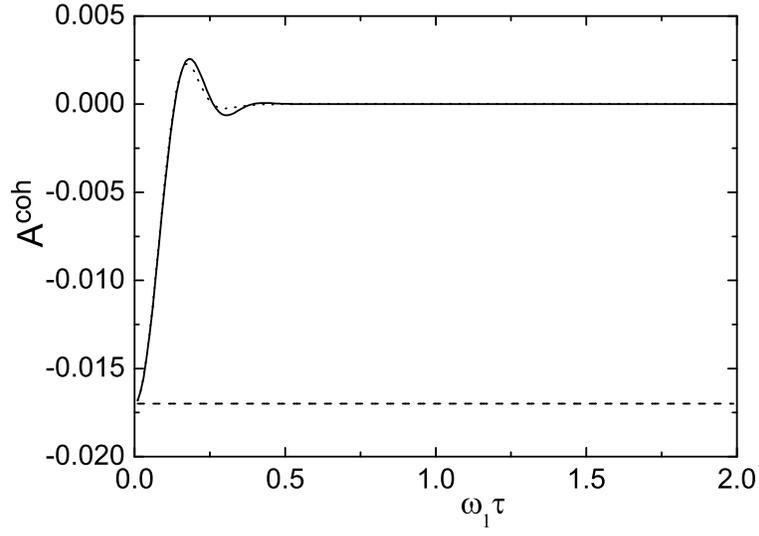}%
\caption{\label{fig:9} Dependence of $A^{coh}$ on $\omega_1\tau$
calculated within semiquasistatic (solid), dynamic (dotted) and
quasistatic (dashed) approaches for $m=5$ and $\Omega_1\tau=2.0$.}
\end{figure}
\begin{figure}
\includegraphics[clip=true,width=0.7\linewidth]{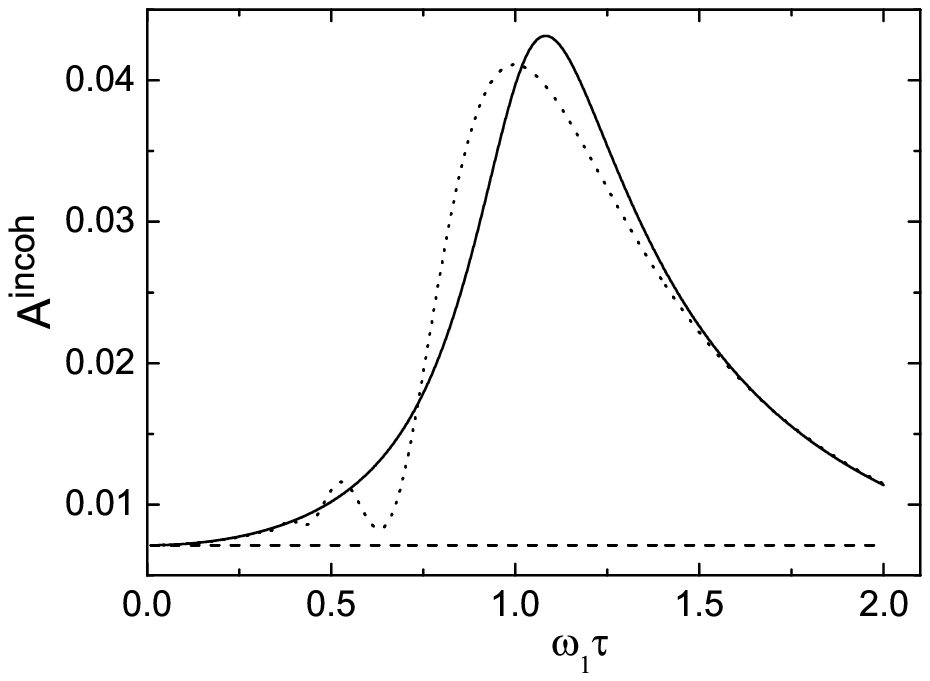}%
\caption{\label{fig:10} Same as in Fig.~\ref{fig:8} but for
$\Omega_1\tau=5.1$.}
\end{figure}
\begin{figure}
\includegraphics[clip=true,width=0.7\linewidth]{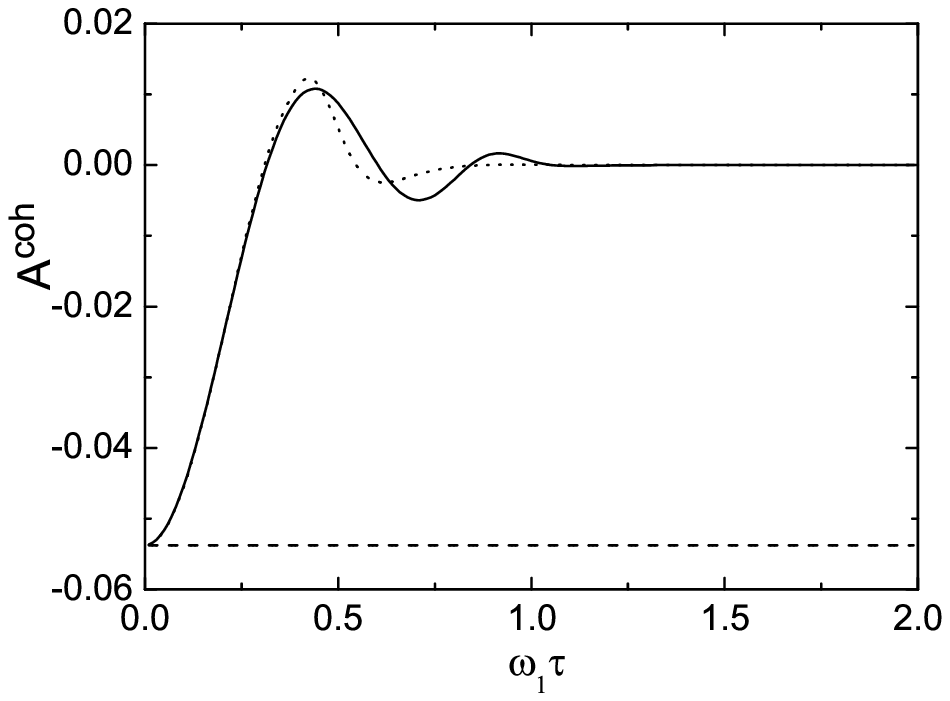}%
\caption{\label{fig:11} Same as in Fig.~\ref{fig:9} but for
$\Omega_1\tau=5.1$.}
\end{figure}

We also calculated numerically $A^{incoh}$ and $A^{incoh}$ using
formulas (\ref{Acoh}), (\ref{Aincoh}) valid for arbitrary values
of $\omega_1\tau$ and $\omega_2\tau$. Results of these
calculations in comparison with semiquasistatic calculations are
shown in Figs.~\ref{fig:8},~\ref{fig:9} for the case of moderate
pump strength and in Figs.~\ref{fig:10},~\ref{fig:11} for a strong
ac pump. For all calculations, including those which are not shown
in these figures, we always found an excellent agreement between
results of dynamic (dots) and semiquasistatic (solid) approaches
for small $\omega_1\tau$. Surprisingly, the semiquasistatic
formulas work quite well even beyond their formal range of
validity, i.e. not only for small values of $\omega_1\tau$. For
the moderate pump, the semiquasistatic approach works well up to
$\omega_1\tau\simeq 2$ (Figures~\ref{fig:8}, \ref{fig:9}). Even
for a strong enough pump, the semiquasistatic approach still gives
a correct smoothed-out curve, while it does not exactly reproduce
small oscillations in the dependence of absorption on the pump
frequency (Figures~\ref{fig:10},\ref{fig:11}).
%=============Figure 8======================================
%%former{fig:7}{fig7.eps}
%===================================================
%=============Figure 9======================================
%%former{fig:6}{fig6.eps}
%===================================================
%=============Figure 10======================================
%%former{fig:9}{fig9.eps}
%===================================================
%=============Figure 11======================================
%%former {fig:8}{fig8.eps}
%===================================================
%
\section{Conclusion}
In summary, we calculated a small-signal absorption (gain) of a
high-frequency field in semiconductor superlattice driven by a
quasistatic ac field.  In particular, our theoretical approach
well describes practically interesting cases of generation and
amplification of THz signal in superlattice devices pumped by a
microwave field. We presented transparent geometrical
interpretations for the coherent and the incoherent contributions
to the net absorption at the different ratio of the pump
($\omega_1$) and the signal ($\omega_2$) frequencies. We
demonstrated that the use of these geometric pictures sufficiently
simplifies analysis: It is easy to find gain for a signal with an
arbitrary $\omega_2$ using the electric characteristics of
superlattice formed under microwave fields. In particular, we
proved that for $\omega_2/\omega_1$ being some irrational number a
small-signal gain at $\omega_2$ is possible only in the condition
of static negative differential conductance. Next, we also found
that an absorption at an arbitrary $\omega_2$, that equals to some
half-integer of the pump frequency, is always zero in an unbiased
superlattice. At least but not last, we presented the universal
analytic procedure for finding the behavior of different
physically important variables in the quasistatic limit starting
from the exact solution of the Boltzmann transport equations for
superlattices. Numerical calculations confirmed our analytic
findings and also demonstrated that the semiquasistatic formulas
for absorption work reasonable well even beyond their formal range
of validity. Our results can be generalized to other physical
systems where the miniband transport regime exist, such as carbon
nanotubes \cite{kibis} or dissipative optical lattices
\cite{grynberg}.
\par
Although in this work we used only a single scattering time, our
results obtained within semiquasistatic approximation can be also
applied to a more realistic case of two different relaxation times
for electron energy and electron momentum. Really, within the
approximation of two relaxation times the static voltage-current
characteristic of superlattice still has the Esaki-Tsu form
\cite{ignatov91}. Because in the semiquasistatic approach only the
(quasi)static voltage-current characteristic plays a role, we
speculate that a variation of the ratio of relaxation times can
lead only to a some change in absolute value of the absorption but
not in its dependence on the amplitude of pump field. Preliminary
numerical simulations of the superlattice balance equations
\cite{ignatov76,ignatov91} confirm these speculations \cite{ahti}.
The detailed consideration of THz absorption with the use of two
different relaxation times will be published elsewhere.
\par
It is instructive to consider our activity from the viewpoint of
its place among other works devoted to the use of the quantum
derivatives or, other words, to the use of the finite differences
in expressions for currents. In the theory of semiconductor
superlattices, the quantum derivative has been first introduced in
\cite{ignatov_q-deriv} for calculations of the absorption of weak
monochromatic ac field in dc-biased superlattice. Later the
formula has been used in the calculations of the response function
of THz superlattice detector \cite{ignatov99} and in the finding
of THz gain in the models of generalized Bloch oscillator with
suppressed domains \cite{wacker97,wackerrew,timo}. Interestingly,
these formulas for superlattices appears to be same as the
formulas describing the response of two-terminal structures to the
action of dc and ac voltages \cite{Tuk79}. The latter are widely
used for calculations of signal mixing in the Josephson junctions
operating in the qusiparticle transport regime and in other
single-barrier tunnel devices \cite{Tuk79,Fel82,Tuk85,Kor94}.
\par
To the best of our knowledge this paper is the first work, where
the quantum derivatives naturally appeared in a description of
microstructure's response to a bichromatic field. It is an
exciting problem to understand whether our semiquasistatic
approach, developed for superlattices operating in the miniband
transport regime, can be generalized to other tunnelling
structures.

\begin{acknowledgments}
This research was supported by Academy of Finland (grants 109758)
and AQDJJ Programme of European Science Foundation. We thank Timo
Hyart for useful discussions and critical reading of the
manuscript.

\end{acknowledgments}
%=======================================================================================
\appendix
\section{\label{app:coherent}Semiquasistatic limit for $A^{coh}$}
In this Appendix we consider the semiquasistatic limit for
$A^{coh}$. We rewrite  $A^{coh}$ (\ref{Acoh}) in terms of $I^{ET}$
(\ref{ET_integr}) as
\begin{eqnarray}
\label{Acohapp1}
A^{coh}&=&\frac{\beta_2}{2}\sum_{l=-\infty}^{\infty}J_{l}(\beta_1)\left[J_{l-2m}(\beta_1)
-J_{l+2m}(\beta_1)\right]I^{ET}(\Omega_0+l\omega_1)
\nonumber\\&=&\frac{\beta_2}{2\tau}\int\limits_{0}^{\infty}\exp\left(-\frac{t}{\tau}\right)\sum_{l=-\infty}^{\infty}J_{l}(\beta_1)
\big[J_{l-2m}(\beta_1)
-J_{l+2m}(\beta_1)\big]\sin\big[(\Omega_0+l\omega_1)t\big]dt.
\end{eqnarray}
After changing indexes, $l\rightarrow-l$, in the first term of
Eq.~(\ref{Acohapp1}) we obtain
\begin{eqnarray}
A^{coh}&=&\frac{\beta_2}{2\tau}\int\limits_{0}^{\infty}\exp\left(-\frac{t}{\tau}\right)
\sum_{l=-\infty}^{\infty}J_{l}(\beta_1)
J_{l+2m}(\beta_1)\Big\{\sin\big[(\Omega_0-l\omega_1)t\big]-\sin\big[(\Omega_0+l\omega_1)t\big]\Big\}\,dt
.
\nonumber\\
\label{Acohapp}
\end{eqnarray}
Using the addition formula for sines we get
\begin{eqnarray}
A^{coh}&=&-\frac{\beta_2}{\tau}\int\limits_{0}^{\infty}\exp\left(-\frac{t}{\tau}\right)
\sum_{l=-\infty}^{\infty}J_{l}(\beta_1)
J_{l+2m}(\beta_1)\sin(l\omega_1t)\cos(\Omega_0t)\,dt.
\nonumber\\
\end{eqnarray}
To make the summation over $l$ we use the identity \cite{Pru90}
\begin{eqnarray}
\label{sum1}
\sum_{l=-\infty}^{\infty}\sin(la)J_l(x)J_{l+m}(y)&=&\sin\left(m\arcsin\frac{x\sin
a}{\sqrt{x^2+y^2-2xy\cos
a}}\right)\nonumber\\&\times&J_m\left(\sqrt{x^2+y^2-2xy\cos
a}\right)
\end{eqnarray}
and get
\begin{eqnarray}
\label{Acohb}
A^{coh}&=&-\frac{\beta_2}{\tau}\int\limits_{0}^{\infty}\exp\left(-\frac{t}{\tau}\right)
\sin\left(2m\arcsin\cos\frac{\omega_1t}{2}\right)J_{2m}\left(2\beta_1\sin\frac{\omega_1t}{2}\right)\cos(\Omega_0t)\,dt.
\nonumber\\
\end{eqnarray}
Now we can take into account that for $\omega_1\tau\ll1$ the main
contribution to the integrals with an exponential factor comes
from the terms formally satisfying $\omega_1t\rightarrow0$. In
this case, we can use the estimations
$$
J_{2m}\left(2\beta_1\sin\frac{\omega_1t}{2}\right)\simeq
J_{2m}\left(\Omega_1t\right),
$$
$$
\sin\left(2m\arcsin\cos\frac{\omega_1t}{2}\right)\simeq(-1)^m\sin(m\omega_1t)=(-1)^m\sin(\omega_2t),
$$
because \cite{Abramovits}
$$
\arcsin(1-z)=\frac{\pi}{2}-\sqrt{2z}, \quad \text{for}\quad
|z|\ll1 .
$$
Substituting these estimations to (\ref{Acohb}) we derive
\begin{eqnarray}
A^{coh}&=&-\frac{(-1)^m\beta_2}{\tau}\int\limits_{0}^{\infty}\exp\left(-\frac{t}{\tau}\right)
\sin(\omega_2t)\cos(\Omega_0t)J_{2m}(\Omega_1t)\,dt \nonumber\\
&=&-\frac{(-1)^m\beta_2}{2\tau}\int\limits_{0}^{\infty}\exp\left(-\frac{t}{\tau}\right)
J_{2m}(\Omega_1t)\big[\sin(\omega_2t-\Omega_0t)+\sin(\omega_2t+\Omega_0t)\big]\,dt.
\end{eqnarray}
Using the integral representation for $J_m(z)$ \cite{Abramovits}
\begin{equation}
\label{Bess}
J_m(z)=\frac{1}{\pi}\int\limits_{0}^{\pi}\cos(z\sin\Theta-m\Theta)\,d\Theta,
\end{equation}
we have
\begin{eqnarray}
\label{coh1}
A^{coh}&=&\frac{(-1)^m}{2\pi\tau}\beta_2\int\limits_{0}^{\pi}\int\limits_{0}^{\infty}\exp\left(-\frac{t}{\tau}\right)
\cos(\Omega_1t\sin\Theta-2m\Theta)\nonumber\\&\times&
\Big\{\sin\big[(\Omega_0-\omega_2)t\big]-\sin\big[(\Omega_0+\omega_2)t\big]\Big\}\,dt\,d\Theta
.
\end{eqnarray}
In this expression we use the formula for a product of sines and
cosines and as a result we get the sum of eight integrals. These
integrals are
$$
\int\limits_{0}^{\pi}\cos\big[\Omega_1t\sin\Theta\pm(\Omega_0-\omega_2)t\big]\sin(2m\Theta)\,d\Theta
=\int\limits_{0}^{\pi}\cos\big[\Omega_1t\sin\Theta\pm(\Omega_0+\omega_2)t\big]\sin(2m\Theta)\,d\Theta=0.
$$
and
\begin{eqnarray}
&&\int\limits_{0}^{\pi} \Big\{\sin\big[\Omega_1t\sin\Theta
+(\Omega_0\mp\omega_2)t\big]-\sin\big[\Omega_1t\sin\Theta
-(\Omega_0\mp\omega_2)t\big]\Big\}
\cos(2m\Theta)\,d\Theta\nonumber\\
&=&2(-1)^m\int\limits_{0}^{\pi}\sin\big[\Omega_1t\cos\Theta
+(\Omega_0\mp\omega_2)t\big]\cos(2m\Theta)\,d\Theta.
\end{eqnarray}
Now (\ref{coh1}) takes the form
\begin{eqnarray}
A^{coh}&=&-\frac{1}{2\pi\tau}\beta_2\int\limits_{0}^{\pi}\int\limits_{0}^{\infty}\exp\left(-\frac{t}{\tau}\right)
\Big\{\sin\big[\Omega_1t\cos\Theta+(\Omega_0-\omega_2)t\big]-
\sin\big[\Omega_1t\cos\Theta+(\Omega_0+\omega_2)t\big]\Big\}\nonumber\\
&\times&\cos(2m\Theta)\,dt\,d\Theta,
\end{eqnarray}
which finally allow us to make the inverse transformation by means
of (\ref{ET_integr}) to the Esaki-Tsu current as
\begin{eqnarray}
A^{coh}&=&\frac{1}{4\pi}\beta_2\int\limits_{0}^{2\pi}\left[I^{ET}(\Omega_1\cos\Theta+\Omega_0+\omega_2)-
I^{ET}(\Omega_1\cos\Theta+\Omega_0-\omega_2)\right]\nonumber\\
&\times&\cos(2m\Theta)d\Theta=\frac{\omega_1}{4\pi}\beta_2\int\limits_{0}^{2\pi/\omega_1}
\left[I^{ET}(\Omega_1\cos(\omega_1t)+\Omega_0+\omega_2)-
I^{ET}(\Omega_1\cos(\omega_1t)+\Omega_0-\omega_2)\right]\nonumber\\
&\times&\cos(2m\omega_1t)dt .
\end{eqnarray}
That is the desirable semiquasistatic representation for
$A^{coh}$.

\section{\label{app:harmonic}Quasistatic limit for $A^{harm}$}

Here we consider the semiquasistatic limit for $A^{harm}$. It is
convenient for our purposes to change indexes $l\rightarrow-l$ in
the first term of (\ref{Aharm}) and rewrite $A^{harm}$ in the
following form
\begin{eqnarray}
&&A^{harm}=\sum_{l=-\infty}^{\infty}J_{l}(\beta_1)J_{l+m}(\beta_1)
\left[(-1)^m\frac{(\Omega_0-l\omega_1)\tau}
{1+(\Omega_0-l\omega_1)^2\tau^2}+\frac{(\Omega_0+l\omega_1)\tau}
{1+(\Omega_0+l\omega_1)^2\tau^2}\right]\nonumber\\
\end{eqnarray}
Now we need to rewrite $A^{harm}$ in terms of $I^{ET}$
(\ref{ET_integr}) as
\begin{eqnarray}
\label{Aharmapp}
&&A^{harm}=\sum_{l=-\infty}^{\infty}J_{l}(\beta_1)J_{l+m}(\beta_1)
\left[(-1)^mI^{ET}(\Omega_0-l\omega_1)+I^{ET}(\Omega_0+l\omega_1)\right]\nonumber\\
&=&\frac{1}{\tau}\int\limits_{0}^{\infty}\exp\left(-\frac{t}{\tau}\right)
\sum_{l=-\infty}^{\infty}J_{l}(\beta_1)J_{l+m}(\beta_1)
\Big\{(-1)^m\sin\big[(\Omega_0-l\omega_1)t\big]+\sin\big[(\Omega_0+l\omega_1)t\big]\Big\}
\,dt\nonumber\\
&=&\frac{1}{\tau}\int\limits_{0}^{\infty}\exp\left(-\frac{t}{\tau}\right)
\sum_{l=-\infty}^{\infty}J_{l}(\beta_1)J_{l+m}(\beta_1)
\Big\{\big[(-1)^m+1\big]\sin(\Omega_0t)\cos(l\omega_1t)\nonumber\\&+&\big[(-1)^m-1\big]\cos(\Omega_0t)
\sin(l\omega_1t)\Big\}\,dt
\end{eqnarray}
We can take the sum over $l$ in (\ref{Aharmapp}) using
Formula~(\ref{sum1}) and the identity  \cite{Pru90}
\begin{eqnarray}
&&\sum_{l=-\infty}^{\infty}\cos(la)J_l(x)J_{l+m}(y)
=\cos\left(m\arcsin\frac{x\sin a}{\sqrt{x^2+y^2-2xy\cos
a}}\right)\nonumber\\&\times&J_m\left(\sqrt{x^2+y^2-2xy\cos
a}\right). \label{sum2}
\end{eqnarray}
As a result we have
\begin{eqnarray}
\label{Aharm1}
A^{harm}&=&\frac{1}{\tau}\int\limits_{0}^{\infty}\exp\left(-\frac{t}{\tau}\right)
\Big\{\big[(-1)^m+1\big]\sin(\Omega_0t)\cos\left(m\arcsin\cos\frac{\omega_1t}{2}\right)
J_m\left(2\beta_1\sin\frac{\omega_1t}{2}\right)
\nonumber\\&+&\big[(-1)^m-1\big]\cos(\Omega_0t)
\sin\left(m\arcsin\cos\frac{\omega_1t}{2}\right)
J_m\left(2\beta_1\sin\frac{\omega_1t}{2}\right)\Big\}\,dt .
\end{eqnarray}
For $\omega_1\tau\ll1$ the main contribution to the integrals with
an exponential factor comes from the terms formally satisfying
$\omega_1t\rightarrow0$. Therefore, the following limits
\begin{eqnarray}
\label{est} \sin\left(m\arcsin\cos\frac{\omega_1t}{2}\right)\simeq
\sin\frac{m\pi}{2},\nonumber\\
\cos\left(m\arcsin\cos\frac{\omega_1t}{2}\right)\simeq
\cos\frac{m\pi}{2},\\
J_m\left(2\beta_1\sin\frac{\omega_1t}{2}\right)\simeq
J_m\left(\Omega_1t\right)\nonumber
\end{eqnarray}
are important. Substituting (\ref{est})  in (\ref{Aharm1}), we
obtain
\begin{eqnarray}
\label{Aharm2}
A^{harm}&=&\frac{1}{\tau}\int\limits_{0}^{\infty}\exp\left(-\frac{t}{\tau}\right)
\left\{\big[(-1)^m+1\big]\sin(\Omega_0t)\cos\frac{m\pi}{2}J_m(\Omega_1t)
\right.\nonumber\\&+&\left.\big[(-1)^m-1\big]\cos(\Omega_0t)\sin\frac{m\pi}{2}
J_m(\Omega_1t)\right\}\,dt .
\end{eqnarray}
Now odd and even values of $m$ must be considered separately.
\par
If $m=2k+1$ $(k=0,1,2,\ldots)$ is an odd number, then
Eq.~(\ref{Aharm2}) becomes
\begin{eqnarray}
\label{Aharm3}
A^{harm}&=&\frac{2(-1)^{k+1}}{\tau}\int\limits_{0}^{\infty}\exp\left(-\frac{t}{\tau}\right)
\cos(\Omega_0t)J_m(\Omega_1t)\,dt
\nonumber\\&=&\frac{2(-1)^{k+1}}{\pi\tau}\int\limits_{0}^{\pi}\int\limits_{0}^{\infty}\exp\left(-\frac{t}{\tau}\right)
\cos(\Omega_0t)\cos(\Omega_1t\sin\Theta-m\Theta)\,dt\,d\Theta,
\end{eqnarray}
where we also used the integral representation (\ref{Bess}) for
$J_m$.
\par
To simplify the expression (\ref{Aharm3}) we will need the
following formula
\begin{eqnarray}
\label{int}
\int\limits_{0}^{\pi}\cos(\Omega_0t)\cos(\Omega_1t\sin\Theta-m\Theta)\,d\Theta
=(-1)^{k+1}\int\limits_{0}^{\pi}\sin[(\Omega_0+\Omega_1\cos\Theta)t]\cos(m\Theta)\,d\Theta.\nonumber\\
\end{eqnarray}
To prove it we first convert the integral as
\begin{eqnarray}
\int\limits_{0}^{\pi}\cos(\Omega_1t\sin\Theta-m\Theta)\,d\Theta=
\int\limits_{0}^{\pi}\sin(\Omega_1t\sin\Theta)\sin(m\Theta)\,d\Theta,
\end{eqnarray}
and then use the formula for a product of sines and cosines
\begin{eqnarray}
2\cos\left(\Omega_0t\right)\sin\left(\Omega_1t\sin\Theta\right)=\sin[(\Omega_0-\Omega_1\sin\Theta)t]+
\sin[(\Omega_0+\Omega_1\cos\Theta)t].
\end{eqnarray}
Taking into account that
\begin{eqnarray}
&&\int\limits_{0}^{\pi}\sin[(\Omega_0-\Omega_1\sin\Theta)t]\sin(m\Theta)\,d\Theta-
\int\limits_{0}^{\pi}\sin[(\Omega_0+\Omega_1\sin\Theta)t]\sin(m\Theta)\,d\Theta\nonumber\\
&=&2(-1)^{k+1}\int\limits_{0}^{\pi}\sin[(\Omega_0+\Omega_1\cos\Theta)t]\cos(m\Theta)\,d\Theta,
\end{eqnarray}
we get (\ref{int}).
\par
Now substituting (\ref{int}) in (\ref{Aharm3}) and making the
inverse transformation to the Esaki-Tsu current, we obtain
\begin{eqnarray}
A^{harm}&=&\frac{2}{\pi\tau}\int\limits_{0}^{\pi}\int\limits_{0}^{\infty}\exp\left(-\frac{t}{\tau}\right)
\sin\big[(\Omega_0+\Omega_1\cos\Theta)t\big]\cos(m\Theta)\,dt\,d\Theta\nonumber\\
&=&\frac{1}{\pi\tau}\int\limits_{0}^{2\pi}I^{ET}(\Omega_0+\Omega_1\cos\Theta)\cos(m\Theta)\,
d\Theta=\frac{\omega_1}{\pi}\int\limits_{0}^{2\pi/\omega_1}I^{ET}
\big(\Omega_0+\Omega_1\cos(\omega_1t)\big)\cos(m\omega_1t)\,dt. \nonumber\\
\end{eqnarray}
\par
We turn now to the consideration of  even harmonics. Combining
(\ref{Aharm2}) for $m=2k$ $(k=1,2,\ldots)$ and the integral
representation (\ref{Bess}), we have
\begin{eqnarray}
\label{Aharm4}
A^{harm}&=&\frac{2(-1)^k}{\tau}\int\limits_{0}^{\infty}\exp\left(-\frac{t}{\tau}\right)
\sin(\Omega_0t)J_m(\Omega_1t)\,dt\nonumber\\
&=&\frac{2(-1)^{k}}{\pi\tau}\int\limits_{0}^{\pi}\int\limits_{0}^{\infty}\exp\left(-\frac{t}{\tau}\right)
\sin(\Omega_0t)\cos(\Omega_1t\sin\Theta-m\Theta)\,dt\,d\Theta.
\end{eqnarray}
Next stage of simplification includes several steps. First, we
take into account that
\begin{eqnarray}
\int\limits_{0}^{\pi}\cos(\Omega_1t\sin\Theta-m\Theta)\,d\Theta
=\int\limits_{0}^{\pi}\cos(\Omega_1t\sin\Theta)\cos(m\Theta)\,d\Theta.
\end{eqnarray}
Second, we apply the formula for a product of sines and cosines.
Finally we also use
\begin{eqnarray}
&&\int\limits_{0}^{\pi}
\Big\{\sin\big[(\Omega_0+\Omega_1\sin\Theta)t\big]+\sin\big[(\Omega_0-\Omega_1\sin\Theta)t\big]\Big\}
\cos(m\Theta)\,d\Theta\nonumber\\
&=&2(-1)^k\int\limits_{0}^{\pi}\sin\big[(\Omega_0+\Omega_1\cos\Theta)t\big]
\cos(m\Theta)\,d\Theta.
\end{eqnarray}
As a result, we can represent (\ref{Aharm4}) in the form suitable
for the inverse transformation to the Esaki-Tsu current and obtain
\begin{eqnarray}
A^{harm}&=&\frac{2}{\pi\tau}\int\limits_{0}^{\pi}\int\limits_{0}^{\infty}\exp\left(-\frac{t}{\tau}\right)
\sin\big[(\Omega_0+\Omega_1\cos\Theta)t\big]
\cos(m\Theta)\,dt\,d\Theta\nonumber\\&=&\frac{1}{\pi}\int\limits_{0}^{2\pi}
I^{ET}(\Omega_0+\Omega_1\cos\Theta)\cos(m\Theta)
d\Theta\nonumber\\
&=&\frac{\omega_1}{\pi}\int\limits_{0}^{2\pi/\omega_1}I^{ET}
\big(\Omega_0+\Omega_1\cos(\omega_1t)\big)\cos(m\omega_1t)\,dt\nonumber
\end{eqnarray}
\par
The last equation  coincides with the corresponding equation for
odd harmonics. Thus,  we conclude that in the quasistatic limit
for both even and odd values of $m$
\begin{eqnarray}
A^{harm}=\frac{\omega_1}{\pi}\int\limits_{0}^{2\pi/\omega_1}I^{ET}
\big(\Omega_0+\Omega_1\cos(\omega_1t)\big)\cos(m\omega_1t)\,dt.\nonumber\\
\end{eqnarray}

%=======================================================================================

\end{document}